\documentclass[aps,10pt,reprint,twocolumn]{revtex4}

\usepackage{epsfig}
\usepackage{graphicx}
\usepackage{dcolumn}
\usepackage{amsmath}
\usepackage{latexsym}
\usepackage{color}
\usepackage{subfig}
\usepackage{mathtools}
\usepackage{footnote}
\usepackage{hyperref}
\hypersetup{
    colorlinks=true,
    linkcolor=blue,
    citecolor=blue,
    filecolor=blue,      
    urlcolor=blue,
    hyperfootnotes=true
}
 
\begin{document}

\title{Novel vortices and the role of complex chemical potential in a rotating holographic superfluid}  

\author{Ankur Srivastav\footnote{
\href{mailto:ankursrivastav@bose.res.in}{ankursrivastav@bose.res.in}}, Sunandan Gangopadhyay\footnote{\href{mailto:sunandan.gangopadhyay@gmail.com}{sunandan.gangopadhyay@gmail.com}\\       \href{mailto:sunandan.gangopadhyay@bose.res.in}{sunandan.gangopadhyay@bose.res.in}}}
\affiliation{Department of Theoretical Sciences, S. N. Bose National Centre for Basic Sciences,\\ Block-JD, Sector-III, Salt Lake City,\\ Kolkata 700106, India}

\begin{abstract}
\noindent In this work, we have analytically devised novel vortex solutions in a rotating holographic superfluid. To achieve this result, we have considered a static disc at the $AdS$ boundary and let the superfluid rotate relative to it. This idea has been numerically exploited in \cite{prdR} where formation of vortices in such a setting was reported. We have found that these vortex solutions are eigenfunctions of angular momentum. We have also shown that vortices with higher winding numbers are associated with higher quantized rotation of the superfluid. We have, then, analysed the equation of motion along bulk AdS direction using St{\"u}rm-Liouville eigenvalue approach. A surprising outcome of our study is that the chemical potential must be purely imaginary. We have then observed that the winding number of the vortices decreases with the increase in the imaginary chemical potential. We conclude from this that imaginary chemical potential leads to less dissipation in such holographic superfluids.
 

\end{abstract}

\maketitle


\section{Introduction}
\noindent The emergence of gauge/gravity duality in the past decade has been instrumental in our understanding on strongly correlated systems. This connection between a $d$-dimensional gravity theory and $(d-1)$-dimensional QFT has been applied to various physical systems ranging from early universe cosmology to condensed matter systems like high-$T_c$ superconductors and strongly coupled superfluids \cite{cph,rgc}, to name a few . The holographic superconductors and superfluids have been studied in various spacetime settings using numerical as well as analytical methods. Some crucial properties associated with these phenomenon have been shown in the past few years \cite{hhh,hhh1,sah,gr1,gr2,sg1,gg1,js,assg,asdgsg,rbsg,gg2,gg3,pmchl}. 
In particular, formation of vortex lattice in holographic superconductors near second critical magnetic field has been shown \cite{mno, prl, cvj1,cvj2, cvj3}. Also, it has been observed that vortices are formed if we rotate a superfluid in a cylindical container. It is known from various experiments that there are a variety of possible vortices in a superfluid under rotation \cite{ovlet, gev}. Existence of such vortices is of prime interest in a holographic superfluid model. Numerical studies leading to the existence of  such vortices in a rotating holographic superfluid have been carried out in \cite{prdR, xtzh}. The study made use of the gauge/gravity duality to  investigate the dynamics of a strongly coupled superfluid in an uniformly rotating disk at a finite temperature. As the angular velocity of the disk is increased above a critical value, a vortex with quantized vorticity gets excited. With further increase of the angular velocity, higher vortices are generated. 
In this paper, we have analytically devised novel vortex solutions for a rotating holographic superfluid model proposed in \cite{prdR}. In our study, we consider that there is a static disc of radius $R$ at the $AdS$ boundary and the superfluid rotates relative to this disc. The superfluid being incompressible, demands no flow along the radial direction and hence it is an equivalent description for the alternate scenario where the superfluid is static in an uniformly rotating disc.
The vortex solutions that we have constructed enjoy circular symmetry in the rotating disc of radius $R$ and each of these solutions are eigenfunctions of the angular momentum. To obtain vortices, we have analysed this model very near to the critical value of rotation $\Omega_{c}$, where superfluid vortex state appears. Remarkably, the rotating superfluid also shows the step transitions of the angular velocity observed in \cite{prdR} leading to the excitation of vortices. 
Interestingly, we have also discovered a linear relation between the winding number associated with these vortices and the angular velocity of the rotating superfluid. \\
It is well established that such a simple holographic superfluid model is parametrized by temperature and chemical potential \cite{hhh1}. If one keeps the temperature fixed then there happens to be a phase transition at a critical value of the chemical potential, $\mu_c$. Above this critical chemical potential, the system is in superfluid phase. We have analysed the equation along bulk AdS direction above $\mu_c$ but very close to it. To solve the holographic system in the bulk direction, we have used a variational technique known as St{\"u}rm-Liouville eigenvalue approach. From this analysis, we observe that the chemical potential must be purely imaginary in order to get a consistent solution. Further, it turns out that there is a decrease in the winding numbers with an increase in the imaginary chemical potential. Appearance of imaginary chemical potential have occured earlier in the literature. A good motivation
to work with imaginary chemical potential arises in non-perturbative studies in quantum chromodynamics (QCD) carried out using techniques of lattice gauge theory. The consequence of an imaginary potential in QCD is the periodicity of the Roberge-Weiss (RW) phase transition \cite{rw}. A holographic understanding of this phase transition has been achieved on an Euclidean spacetime set up in \cite{gasp,jr}. In our study, however, the need for an imaginary chemical potential arises in a geometry whose signature is Lorentzian. It
would therefore be interesting to see whether the results in the Euclidean set up would still be applicable when the signature of the spacetime geometry is Lorentzian. At present we can only say it will since the temporal component of the gauge field vanishes at the black hole horizon as it happens in the Euclidean scenario. This would
require further investigation which we shall not carry out here.\\
In order to understand imaginary chemical potential in this holographic model, we have further solved the time-dependent equations for the matter field. We have been able to show that in this timedependent case, imaginary chemical potential competes with the imaginary frequency, which is related to the dissipation in the system. Hence, we conclude that increase in the imaginary chemical potential leads to decrease in vortex number, which implies less dissipation in the system. It should be noted that the complex chemical potential has been related to dissipation in \cite{ck}. Also, a holographic model for the color superconductivity
in QCD with imaginary chemical potential was studied recently \cite{kg}.

\noindent We have organized this paper in the following way. In section (\hyperlink{sec2}{II}), we set up the model for holographic superfluid in a static black hole background in $AdS_{3+1}$ spacetime. In section (\hyperlink{sec3}{III}), we have constructed vortex solutions, near critical rotation in the rotating disc. Section (\hyperlink{sec4}{IV}) deals with the St{\"u}rm-Liouville eigenvalue analysis. Then in the last section (\hyperlink{sec5}{V}) of this paper, we have concluded and made some remarks on our results. 
 
 


\section{The Holographic Superfluid}
\noindent \hypertarget{sec2}{We} start by writing down the metric for a static black hole in $AdS_{3+1}$ spacetime with Eddington-Finkelstein coordinates \cite{prdR},
\begin{eqnarray}
 ds^{2}=\dfrac{\textit{l}^2}{u^2}[-f(u)dt^{2}-2dtdu+dr^{2}+r^{2}d\theta^{2}]
\label{metric}
\end{eqnarray}
where the blackening factor is given by, $$f(u)=\bigg(1-u^3\bigg)~.~$$ Here $\textit{l}$ is the $AdS$ radius and $u$ is the bulk direction scaled in such a way that $u=0$ is the AdS boundary and $u=1$ is the event horizon of the black hole. The coordinates $(r, \theta)$ define the $2D$ flat disc. For convenience, 
we take unit $AdS$ radius (that is, $l = 1$) and the cosmological constant $\Lambda=-3$. The Hawking temperature associated with the above black hole geometry is given by $T=\dfrac{3}{4\pi}$.

\noindent We now consider a simple model for holographic superfluid on top of this geometry. The action for the matter section in this model is given by,
 \begin{eqnarray}
~~~~~~~~~ \mathcal{S}= \dfrac{l^2}{2\kappa_4^2 e^2} \int_\mathcal{M} d^4x ~\mathcal{L}_m ~.
\label{Matter_Action}
\end{eqnarray}
The matter Lagrangian density, $\mathcal{L}_m$, consists of a Maxwell field and a complex scalar field minimally coupled to $A_{\mu}$. More precisely $\mathcal{L}_m$ is given by following expression,
\begin{eqnarray}
~~~~~~~~~ \mathcal{L}_m=-\dfrac{1}{4}F^{\mu\nu}F_{\mu\nu}-|D\Psi|^{2}-m^{2}|\Psi|^{2}
\label{Lagrangian_density}
\end{eqnarray}
$$F_{\mu\nu} \equiv \partial_{[\mu}A_{\nu]} ~ , ~ D \equiv (\nabla-ieA)~$$
where $m$ is the mass of the scalar field while $e$ is its charge. Note that we will be working in the probe limit. In this limit any backreaction of the matter field in the metric is neglected. To achieve this limit, we shall rescale $A_\mu \rightarrow \dfrac{A_\mu}{e}$ and $\Psi \rightarrow \dfrac{\Psi}{e}$ and take the limit $e \rightarrow \infty$. Mathematically, it is equivalent to setting $e = 1$ in the action of our theory. \\
Now varying the action, $\mathcal{S}$, for $\Psi$ and $A_\mu$ we get the following equations of motion for the matter and the gauge fields repectively,
\begin{eqnarray}
    (D^2 - m^2)\Psi = 0 
\label{matter_field_eq}
\end{eqnarray}
\begin{eqnarray}
   \nabla_\nu F_{\mu}^{~\nu} = j_{\mu} 
\label{gauge_field_eq}
\end{eqnarray}
where the bulk current is defined as,  
\begin{eqnarray}
j_{\mu} \coloneqq i\{(D_{\mu}\Psi)^{\dagger} \Psi - \Psi (D_{\mu}\Psi)\} ~.
\label{bulk_current}
\end{eqnarray}
\noindent We shall now assume that all the fields are stationary as our interest lies in equilibrium analysis of the rotating superfluid system. Also we would be working with the axial gauge, that is, $A_u = 0$, in which case eq.(\ref{matter_field_eq}) reduces to the following equation,
\begin{eqnarray}
\{\mathcal{D}(u) + \mathcal{D}(r) + \dfrac{1}{r^2}\mathcal{D}(\theta)\}\Psi(u,r,\theta) = 0
\label{matter_field_eq_in_axial_gauge}
\end{eqnarray}
where the segregated derivative operators are given as,
\begin{eqnarray}
\nonumber
\mathcal{D}(u) \equiv u^2\partial_u\Big(\dfrac{f(u)}{u^2}\partial_u\Big) + i u^2 \partial_u\Big(\dfrac{A_t}{u^2}\Big) + iA_t \partial_u - \dfrac{m^2}{u^2} \\ 
\nonumber
\mathcal{D}(r) \equiv \dfrac{1}{r}\partial_r(r\partial_r) - \dfrac{i}{r} \partial_r(r A_r) - iA_r \partial_r - A_r^2 ~~~~~~~~~~~\\
\nonumber
\mathcal{D}(\theta)  \equiv \partial_\theta^{~2} - i(\partial_\theta A_\theta + A_\theta \partial_\theta) - A_\theta ^{~2}~.~~~~~~~~~~~~~~~~~~~
\label{Diffrential_Operators}
\end{eqnarray}

\section{The Vortex Solution}
\noindent \hypertarget{sec3}{Our} interest is in the equilibrium state where vortices exist. So we  define a deviation parameter, $\epsilon$, from the critical rotation, $\Omega_{c}$, by the following relation, 
\begin{eqnarray}
\epsilon \coloneqq \dfrac{\Omega - \Omega_{c}}{\Omega_{c}} 
\label{deviation_parameter}
\end{eqnarray}
where $\Omega$ is the constant angular velocity of the disc. As argued in \cite{prdR}, one should notice that there is a relative velocity between the superfluid and the disc. Hence, a static superfluid in a rotating disc is justly represented by a rotating superfluid in a static disc. In this analysis, we are visualizing the latter scenario. 
Now, in order to study this system very near to $\Omega_{c}$, we series expand the matter field $\Psi$, the gauge field $A_\mu$ and the bulk current $j_\mu$ with respect to $\epsilon$ in the following manner \cite{mno},
\begin{eqnarray}
\Psi(u,r,\theta) = \sqrt{\epsilon}\Big(\Psi_1(u,r,\theta) + \epsilon \Psi_2(u,r,\theta)+...\Big)\\ 
A_\mu(u,r,\theta) = \Big(A_\mu^{(0)}(u,r,\theta) + \epsilon A_\mu^{(1)}(u,r,\theta)+...\Big)\\
j_\mu(u,r,\theta) = \epsilon \Big(j_\mu^{(0)}(u,r,\theta) + \epsilon j_\mu^{(1)}(u,r,\theta)+...\Big)~.
\label{Series expansion in epsilon}
\end{eqnarray}



\subsection{Zeroth order solutions near AdS boundary}

\noindent The zeroth order solutions for gauge fields, in axial gauge, that generates the critical rotation field and the chemical potential are given by following relations,
\begin{eqnarray}
 A_t^{(0)}(u) = \mu(1-u), ~~~ A_r^{(0)} = 0, ~~~ A_\theta^{(0)}(r) = \Omega r^2 ~.
\label{gauge fields in zeroth order}
\end{eqnarray}
Notice that $A_r^{(0)} = 0$ restricts any superfluid flow in the radial direction while $A_\theta^{(0)}$ allows the superfluid to rotate.\\ 
Considering these zeroth order solutions for gauge fields near the $AdS$ boundary, we may rewrite eq.(\ref{matter_field_eq_in_axial_gauge}) for lowest order in $\epsilon$, that is, $\mathcal{O}(\sqrt{\epsilon})$, in the following form,
\begin{eqnarray}
\{\mathcal{D}^{(0)}(u) + \mathcal{D}^{(0)}(r) + \dfrac{1}{r^2}\mathcal{D}^{(0)}(\theta)\}\Psi_1(u,r,\theta) = 0
\label{matter_field_eq_lowest_epsilon_order}
\end{eqnarray}
where the derivative operators become,
\begin{eqnarray}
\nonumber
\mathcal{D}^{(0)}(u) \equiv u^2\partial_u\Big(\dfrac{f(u)}{u^2}\partial_u\Big) + i u^2 \partial_u\Big(\dfrac{A_t^{(0)}}{u^2}\Big) + iA_t^{(0)} \partial_u - \dfrac{m^2}{u^2} \\ 
\nonumber
\mathcal{D}^{(0)}(r) \equiv \dfrac{1}{r}\partial_r(r\partial_r)  ~~~~~~~~~~~~~~~~~~~~~~~~~~~~~~~~~~~~~~~~~~~~~~~~~~\\
\nonumber
\mathcal{D}^{(0)}(\theta)  \equiv \partial_\theta^{~2} - i(\partial_\theta A_\theta^{(0)} + A_\theta^{(0)} \partial_\theta) - A_\theta ^{(0)2} ~.~~~~~~~~~~~~~~~~~~
\label{zeroth_order_Diffrential_Operators}
\end{eqnarray}
We now use the method of separation of variables 
to solve eq.(\ref{matter_field_eq_lowest_epsilon_order}) 
and write $\Psi_1(u,r,\theta)$ as a function of $u$ and $(r,\theta)$ 
separately in the following manner,
\begin{eqnarray}
\Psi_1(u,r,\theta) = \Phi(u) \xi(r,\theta)~.
\label{first_separation}
\end{eqnarray}
With the above separation of matter field, eq.(\ref{matter_field_eq_lowest_epsilon_order}) provides the following separated equations,
\begin{eqnarray}
\mathcal{D}^{(0)}(u)\Phi(u) = \lambda \Phi(u)  
\label{matter_field_eq_in Ads direction}
\end{eqnarray}
\begin{eqnarray}
\{\mathcal{D}^{(0)}(r)+\dfrac{1}{r^2} \mathcal{D}^{(0)}(\theta)\}\xi(r,\theta) = - \lambda \xi(r,\theta)
\label{matter_field_eq_in disc}
\end{eqnarray}
where $\lambda$ is an unknown separation constant. 
Note that both 
eq.(s)(\ref{matter_field_eq_in Ads direction}, \ref{matter_field_eq_in disc})
are eigenvalue equations with eigenvalue $\lambda$. In the subsequent discussion we shall proceed to determine $\lambda$.



\subsection{Solution for vortex in the rotating superfluid}

\noindent Given the $2D$ rotational symmetry, we may choose the following ansatz,
\begin{eqnarray}
\xi(r,\theta) = \eta_p(r) e^{ip\theta}
\label{r_theta_separation}
\end{eqnarray}
where $p \in \mathcal{Z}$ for the single valuedness of the solution. However, one should note that $\eta_p(r)$ must satisfy certain boundary conditions for regularity at the boundaries. In our case, we would be working with the Neumann boundary conditions at $r = 0$ as well as at $r = R$, that is,
\begin{eqnarray}
\partial_r\eta_p|_{r=0} = 0 = \partial_r\eta_p|_{r=R} 
\label{boundary_conditions}
\end{eqnarray}
where $R$ is the radius of the disc boundary. \\ 
Now using the above ansatz in eq.(\ref{matter_field_eq_in disc}), we get the following differential equation to be solved under the boundary conditions defined above,
\begin{eqnarray}
\partial_r^{~2}\eta_p(r) + \dfrac{1}{r} \partial_r\eta_p(r) + \Big\{\lambda - \Big(\dfrac{p}{r} - \Omega r\Big)^2\Big\}\eta_p(r) = 0.
\label{matter eq in r direction}
\end{eqnarray}
To solve for $\eta_p(r)$, we consider the following ansatz, 
\begin{eqnarray}
\eta_p(r) = F_p(r) e^{-\Omega r^2/2}~.
\label{eta_p solution ansatz}
\end{eqnarray}
Utilising this form of $\eta_p(r)$ given by eq.(\ref{eta_p solution ansatz}), 
eq.(\ref{matter eq in r direction}) takes the form, 
\begin{equation}
\partial_r^{~2}F_p(r) + \Big(\dfrac{1}{r} - 2p\Omega \Big) \partial_rF_p(r)   +\Big(\tilde{\lambda} -2\Omega - \dfrac{p^2}{r^2}\Big)F_p(r) = 0.
\label{matter eq reduced in r direction}
\end{equation}
We now proceed to solve eq.(\ref{matter eq reduced in r direction}) using the 
Frobenius series solution method. So we consider that $F_p(r)$ is given by the following series, 
\begin{eqnarray}
F_p(r) =  \sum_{n=0}^{\infty} a_n r^{n+k} ~~~~,~~~~ (a_0 \neq 0)
\label{series_solution}
\end{eqnarray}
with $k$ being an integer.
The derivatives of the above series solution with respect to $r$ are given by,
\begin{eqnarray}
\partial_r F_p(r) = \sum_{n=0}^{\infty} a_n (n+k) r^{n+k-1}
\label{series_solution_first_deri}
\end{eqnarray}
\vspace*{-7mm}

and
\vspace*{-7mm}

\begin{eqnarray}
 \partial_r^2 F_p(r) = \sum_{n=0}^{\infty} a_n (n+k)(n+k-1) r^{n+k-2}~.
\label{series_solution_second_deri}
\end{eqnarray}
Using eqs.(\ref{series_solution}, \ref{series_solution_first_deri} and \ref{series_solution_second_deri}) in eq.(\ref{matter eq reduced in r direction}), we find the following condition, 
\begin{eqnarray}
\nonumber
\sum_{n=0}^{\infty} a_n \{(n+k)^2 - p^2)\} r^{n+k} +\\
 \sum_{n=0}^{\infty} a_n \{\lambda + 2\Omega(p-1-n-k)\}r^{n+k+2}= 0~.
 \label{coefficient equation}
\end{eqnarray}
This implies that coefficient for each order of $r$ should separately satisfy eq.(\ref{coefficient equation}), that is, 
\begin{eqnarray}
\nonumber
r^k~~~ : ~~~~~ a_0 (k^2 - p^2) = 0 ~~\Longrightarrow ~~k = \pm p ~~~~~~~~~~~~~~~ \\
\nonumber
r^{k+1} : ~~~~~ a_1 ((k+1)^2 - p^2) = 0 ~~\Longrightarrow ~~ (k+1) = \pm p ~.
 \label{coefficient equation for lowest order}
\end{eqnarray}
From the above conditions we consider $k = p$ for the regularity of the solutions at $r=0$ and this yields $a_1 = 0$. The condition $k=p$ implies that $p$ is an integer. Similarly setting the coefficient for $r^{(k+n+2)}$ equal to zero, we get the following recurrence relation,
\begin{eqnarray}
 \dfrac{a_{n+2}}{a_n} =  \dfrac{(\lambda - 2\Omega (n+1))}{((n+2)^2 + 2p(n+2))} 
 \label{recurrence_relation}
\end{eqnarray}
where we have already used the condition $k=p$. This recurrence relation connects all the even coefficients with $a_0$ and all the odd coefficients with $a_1$. Hence, we would get series solution for $F_p(r)$ with even terms only.
Now in order to have normalizable solutions, we must terminate this series at some point, which determines $\lambda$ in terms of $\Omega$ and $n$, that is, 
\begin{eqnarray}
\lambda = 2\Omega (n+1).
\label{eigenvalue}
\end{eqnarray}
The above relation implies that the eigenvalue $\lambda$ is quantized.
With this condition, the above series solution becomes a polynomial of order $n$. Thus we can write the solution for $\eta_p (r)$ with an additional index depicting the order of the polynomial as, 
\begin{eqnarray}
\eta_{p,n}(r) = a_0 e^{-\Omega r^2/2} F_{p,n}(r)
\label{final_eta_solution}
\end{eqnarray}
where $$F_{p,n}(r) = r^p \big(1 + \dfrac{a_2}{a_0} r^2 + \dfrac{a_4}{a_0} r^4 + ... + \dfrac{a_n}{a_0} r^n\big)~.$$
Let us now discuss the family of solutions with $n=0$. In this case, $$F_{p,0}(r) = r^p$$  and hence, 
\begin{eqnarray}
\eta_{p,0}(r) = a_0 r^p e^{-\Omega r^2/2}~~~~;~~~(\lambda = 2\Omega)
\label{final_eta_solution_lowest_order}
\end{eqnarray}
This solution is subjected to the Neumann boundary conditions mentioned earlier. This means the following first derivative of eq.(\ref{final_eta_solution_lowest_order}) must vanish at the disc boundaries, 
\begin{eqnarray}
 \partial_r\eta_{p,0}(r) = a_0 r^{p-1} e^{-\Omega r^2/2} (p - \Omega r^2)~. 
 \label{first_derivative}
\end{eqnarray}
Now the boundary condition at $r=0$ gives the following lower bound for $p$,
\begin{eqnarray}
 \partial_r\eta_{p,0}(r)|_{r=0} = 0 ~~~~\Longrightarrow ~~~~p > 1.
 \label{first_restriction_on_p}
\end{eqnarray}
Applying the boundary condition at the disc boundary at $r=R$ gives the following linear relation between $p$ and $\Omega$,  
\begin{eqnarray}
\partial_r\eta_{p,0}(r)|_{r=R} = 0 ~~~~\Longrightarrow ~~~~ p = \Omega R^2.
\label{second_restriction_on_p}
\end{eqnarray}
Since $p$ is an integer, hence the above relation between $p$ and $\Omega$ implies a quantization of the angular velocity $\Omega$ and also a quantization of the angular momenta in the rotating superfulid. Note that the radius $R$ in the model is fixed. This implies that there is a linear relation between $p$ and $\Omega$. This is a nice result that comes from our analysis.\\
\noindent Let us now consider the solution for $n=2$, which is given as,
\begin{eqnarray}
 \eta_{p,2}(r) = a_0 r^p e^{-\Omega r^2/2}\Big(1 - \dfrac{2\Omega}{(p+2)}r^2\Big)~~;~(\lambda = 6\Omega)~.
 \label{final_eta_solution_first_order}
\end{eqnarray}
For this solution, we have, 
\begin{eqnarray}
\partial_r\eta_{p,2}(r) = a_0 r^{p-1} e^{-\Omega r^2/2} \Big(p - 3\Omega r^2 + \dfrac{2 (\Omega r^2)^2}{p+2}\Big).
\label{first_derivative1}
\end{eqnarray}
In this case, the boundary condition at $r=0$ gives us the same lower bound for $p$,
\begin{eqnarray}
\partial_r\eta_{p,2}(r)|_{r=0} = 0 ~~~~\Longrightarrow ~~~~p > 1.
\label{first_restriction_on_p1}
\end{eqnarray}
However, the boundary condition at $r=R$ gives us the following condition,
\begin{eqnarray}
\partial_r\eta_{p,2}(r)|_{r=R} = 0 ~\Longrightarrow ~ \Big(p - 3\Omega R^2 + \dfrac{2 (\Omega R^2)^2}{p+2}\Big) = 0.~~
\label{second_restriction_on_p1}
\end{eqnarray}
From this condition, we get, 
\begin{eqnarray}
\Omega R^2 = 3\dfrac{(p+2)}{4}\Big(1\pm\sqrt{1-\dfrac{8p}{9(p+2)}}\Big).
\label{second_restriction_on_p2}
\end{eqnarray}
For $p>>2$, the above result again provides a linear relation between $p$ and $\Omega$, that is, $\Omega R^2 \sim p$.

\begin{figure}
\includegraphics[width = \linewidth]{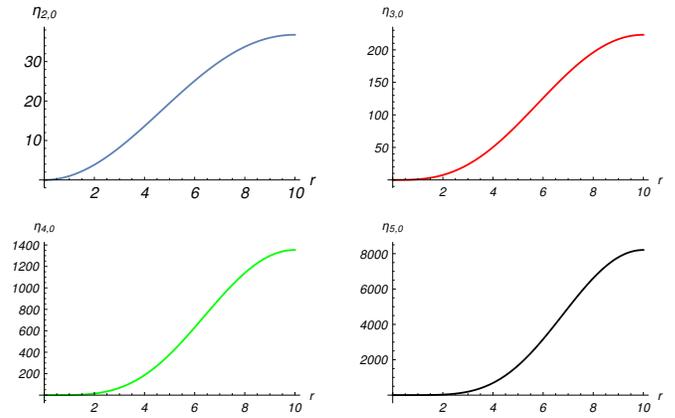}
\caption{Un-normalized lowest order ($n=0$) vortex solutions for different winding numbers. (The value of R is set to be equal to 10).}
\end{figure}
\section{St{\"u}rm-Liouville Eigenvalue Analysis}
\noindent In this section, we shall solve eq.(\ref{matter_field_eq_in Ads direction}) using St{\"u}rm-Liouville eigenvalue approach. We shall consider the analysis near the critical chemical potential ($\mu \sim \mu_c$) so that we may take the following ansatz for the gauge fields near the AdS boundary \footnote{$A_t^{(0)}(u) = \mu (1-u)\simeq \mu $ for $u \rightarrow 0$. Note that $A_t^{(0)}(u)$ vanishes at the black hole horizon $u = 1$.},
\begin{eqnarray}
 A_t^{(0)}(u) = \mu, ~~~ A_r^{(0)} = 0, ~~~ A_\theta^{(0)}(r) = \Omega r^2 ~.
\label{gauge fields approx in zeroth order}
\end{eqnarray}
For simplicity, we shall consider $m^2 = -2, \Delta = 1$. With these considerations, eq.(\ref{matter_field_eq_in Ads direction}) reduces to the following equation,
\begin{eqnarray}
 u^2 \partial_u \Big(\dfrac{1-u^3}{u^2} \partial_u \Phi(u) \Big) + i u^2 \partial_u \Big(\dfrac{\mu}{u^2} \Phi(u) \Big) \nonumber \\
 + i \mu \partial_u \Phi(u) + \dfrac{2}{u^2} \phi = 2\Omega \Phi(u) ~.
\label{bulk equation 01}
\end{eqnarray}
Notice that we have considered only the case for $n=0$ and hence $\lambda = 2\Omega$. We now simplify eq.(\ref{bulk equation 01}) in the following form,
\begin{eqnarray}
 (1-u^3)\partial_u^{2}\Phi-\Big(u^2+\dfrac{2}{u}-2i\mu\Big)\partial_u \Phi  \nonumber \\-\Big(2\Omega-\dfrac{2}{u^2}+\dfrac{2i\mu}{u}\Big) \Phi = 0~. 
\label{bulk equation 02}
\end{eqnarray}
Near AdS boundary ($u \rightarrow 0$), we can write $\Phi(u)$ in the following manner, $$\Phi(u) \simeq~ < \mathcal{O}_1> u \Lambda(u) $$ so that $\Lambda(u)$ is subjected to the boundary conditions given below,
\begin{equation}
\Lambda(0) = 1 ~~;~~ \partial_u\Lambda(0)=0~.
\label{bdy cond for Lambda}
\end{equation}
Using this in eq.(\ref{bulk equation 02}), we get an equation for $\Lambda$ as given below,
\begin{eqnarray}
(1-u^3) \Lambda^{\prime\prime} - (3u^2-2i\mu)\Lambda^{\prime} -(u+2\Omega)\Lambda = 0~
\label{Lambda eq 01}
\end{eqnarray}
where $^{\prime}$ denotes derivative with respect to $u$. Considering $\Lambda$ to be real, eq.(\ref{Lambda eq 01}) implies that $\mu$ must be purely imaginary for eq.(\ref{Lambda eq 01}) to have a consistent solution. So we have $Re(\mu) = 0$, and set $Im(\mu) = \mu^I$. With this, eq.(\ref{Lambda eq 01}) becomes,
\begin{eqnarray}
(1-u^3) \Lambda^{\prime\prime} - (3u^2+2\mu^I)\Lambda^{\prime} -(u+2\Omega)\Lambda = 0~.
\label{Lambda eq 02}
\end{eqnarray}
In order to cast eq.(\ref{Lambda eq 02}) into St{\"u}rm-Liouville form, we multiply it with integrating factor $R(u)$ given below,
\begin{eqnarray}
R(u) = \Big( \dfrac{1-u}{\sqrt{1+u+u^2}}\Big)^{\dfrac{2\mu^I}{3}} exp\Big(-\dfrac{2\mu^I}{\sqrt{3}} \arctan(\dfrac{1+2u}{\sqrt{3}})\Big)~.\nonumber 
\label{IF}
\end{eqnarray}
With this, eq.(\ref{Lambda eq 02}) can be put into St{\"u}rm-Liouville form as given below,
\begin{eqnarray}
(P(u)\Lambda^{\prime}(u))^{\prime}+ Q(u)\Lambda(u) + \Omega S(u)\Lambda (u) =0~ 
\label{SL equation}
\end{eqnarray}
where  
\begin{eqnarray}
P(u) = (1-u^3) R(u) \nonumber \\  Q(u) = -u R(u)~~~~~~\nonumber \\
S(u) = 2R(u)~.~~~~~~~\nonumber
\label{SL Coefficients}
\end{eqnarray}
Now the eigenvalue $\Omega$ is given by the following integral,
\begin{eqnarray}
\Omega = \dfrac{\int_0^1 du (P(u)(\Lambda^{\prime}(u))^2-Q(u)\Lambda^{2} (u))}{\int_0^1 du S(u)\Lambda^{2} (u) }~.
\label{SL eigen01}
\end{eqnarray}
In order to proceed ahead, we take a trial function for $\Lambda(u)$ that satisfies the given boundary conditions, that is, $\Lambda(0) = 1 ,~ \partial_u\Lambda(0)=0~.$ We assume the following trial function, $$\Lambda_{\alpha}(u) = (1-\alpha u^2)~.$$
With this trial function, we have the following equation to determine $\Omega_{\alpha}$,
\begin{eqnarray}
\Omega_{\alpha} = \dfrac{\int_0^1 du (P(u)(\Lambda_{\alpha}^{\prime}(u))^2-Q(u)\Lambda_{\alpha}^{2} (u))}{\int_0^1 du S(u)\Lambda_{\alpha}^{2} (u) }~.
\label{SL eigen02}
\end{eqnarray}
In order to compute eq.(\ref{SL eigen02}), we approximate $R(u)$ for $u \rightarrow 0$ in the following manner,
\begin{eqnarray}
R(u) \simeq \Big(1-\dfrac{2\mu^I}{\sqrt{3}}\arctan(\dfrac{1+2u}{\sqrt{3}})\Big)~.
\label{IF approx}
\end{eqnarray}
Using eq.(\ref{IF approx}) into eq.(\ref{SL eigen02}), we find the following equation for $\Omega_{\alpha}$, 
\begin{eqnarray}
\Omega_{\alpha} = ~~~~~~~~~~~~~~~~~~~~~~~~~~~~~~~~~~~~~~~~~~~~~~~~~~~~~~~~~~~~~~~~~~~~~~~~~ \label{SL eigen03}\\ \dfrac{\int_0^1 du \Big(1-\dfrac{2\mu^I}{\sqrt{3}} \arctan(\dfrac{1+2u}{\sqrt{3}})\Big)\big(u+4\alpha^2 u^2 -2\alpha u^3 - 3\alpha^2 u^5\big)}{\int_0^1 du \Big(1-\dfrac{2\mu^I}{\sqrt{3}} \arctan(\dfrac{1+2u}{\sqrt{3}})\Big)\big(1+\alpha^2 u^4 -2\alpha u^2\big)}~.\nonumber 
\end{eqnarray}
We now need to extremize $\Omega_{\alpha}$ with respect to $\alpha$. For a fix value of $\mu^I$, it turns out that there are two values of $\alpha$ which extremize eq.(\ref{SL eigen03}). To understand the qualitative role of $\mu^I$, we have calculated these extremized values of $\Omega_{\alpha}$ for a range of values of $\mu^I$. Extremized values of $\Omega_{\alpha}$, corresponding to both values of $\alpha$, for $\mu^I$ between $4.0$ and $4.5$ are shown in Fig.(2) and Fig.(3).
\begin{figure}
\includegraphics[width = \linewidth]{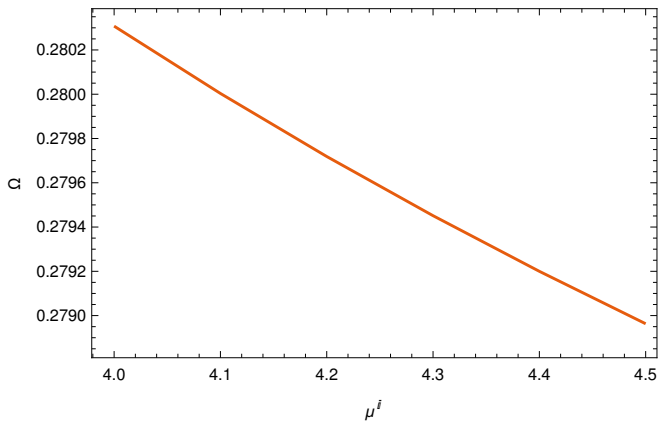}
\caption{$\Omega$ vs $\mu^I$ for lowest order ($n=0$) vortex solutions for first values of $\alpha$ that extremize $\Omega_{\alpha}$ in eigenvalue equation (\ref{SL eigen03}).}
\end{figure}
\begin{figure}
\includegraphics[width = \linewidth]{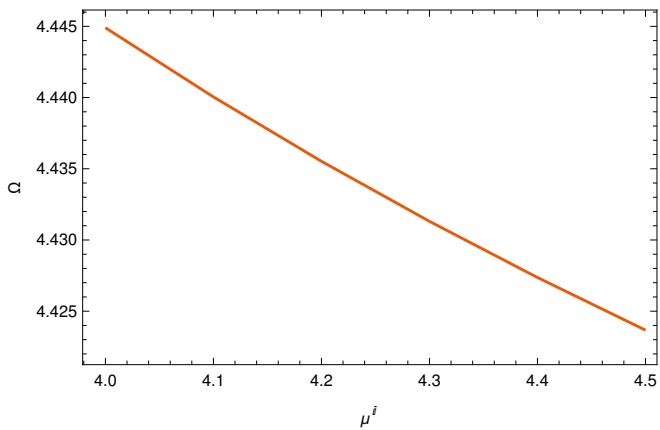}
\caption{$\Omega$ vs $\mu^I$ for lowest order ($n=0$) vortex solutions for second values of $\alpha$ that extremize $\Omega_{\alpha}$ in eigenvalue equation (\ref{SL eigen03}).}
\end{figure}
These figures show a remarkable trend, in both the cases extremized values of $\Omega$ consistently decreases with an increase in the value of imaginary chemical potential. Now some subtle observations are in order here. As we have shown in section (\hyperlink{sec3}{III}) that these $\Omega$ are quantized with the following relation, $$\Omega = \dfrac{p}{R^2}~$$ where $R$ is the radius of the disc. This relation in conjuction with Fig.(2) and Fig.(3) implies that for a disc with a fixed radius $R$, there is a decrease in the winding numbers as the imaginary chemical potential rises. This seems to be an interesting observation from holographic point of view. In order to better understand this result, we have further considered the timedependent terms in the equation of motions given by eq.(\ref{matter_field_eq_in_axial_gauge}). The corresponding time-dependent equation is given as,
\begin{eqnarray}
\{\mathcal{D}(u) + \mathcal{D}(r) + \dfrac{1}{r^2}\mathcal{D}(\theta)-2\partial_u\partial_t + \dfrac{2}{u}\partial_t\}\Psi(t,u,r,\theta) = 0.~~
\label{time-dependent matter_field_eq_in_axial_gauge}
\end{eqnarray}
Linearizing this equation with the following form of $\delta\Psi$ and $\delta A_{\mu}$ along with the boundary conditions expressed before, 
$$\delta\Psi = p(u,r) e^{i\omega t+in\theta}~~;~~\delta A_{\mu} = a_t(u,r) e^{i\omega t+in\theta}$$
gives the following equations after separation of variables for $p(u,r) = \Phi(u)\eta_p(r)$,
\begin{eqnarray}
 (1-u^3)\partial_u^{2}\Phi-\Big(u^2+\dfrac{2}{u}-2i(\mu-\omega)\Big)\partial_u \Phi  \nonumber \\-\Big(-\dfrac{2}{u^2}+\dfrac{2i(\mu-\omega)}{u}\Big) \Phi = \lambda \Phi ~.
\label{time dep. eom}
\end{eqnarray}
\begin{eqnarray}
\partial_r^{~2}\eta_p(r) + \dfrac{1}{r} \partial_r\eta_p(r) + \Big\{\lambda - \Big(\dfrac{p}{r} - \Omega r\Big)^2\Big\}\eta_p(r) = 0.
\label{matter eq in r direction 02}
\end{eqnarray}
One should notice that these are similar equations that we have found for stationary field case in sections (\hyperlink{sec3}{III}) and (\hyperlink{sec4}{IV}). The only difference is that in eq.(\ref{time dep. eom}), $\mu$ is now replaced with $(\mu - \omega)$. This difference immediately points towards a connection between imaginary chemical potential and imaginary part of the frequency, $\omega$. As it is well known that imaginary part of the frequency, $\omega$, implies dissipation in the system, hence we may attach a similar meaning to $\mu^I$. We observe from Fig.(2) and Fig.(3) that there is a decrease in the number of vortices with a rise in the value of the imaginary chemical potential. On the other hand, from Fig.(4), we observe that with increase in $\omega$, $\Omega$ increases which means that the number of vortices increases. Now increase in the vortex number can be understood as an increase of dissipation in the system \cite{pmchl}. Hence, the presence of both the imaginary chemical potential,`$\mu^i$' and the frequency `$\omega$' of the quasi-normal modes reduces the dissipation in the system. 

\begin{figure}
\includegraphics[width = \linewidth]{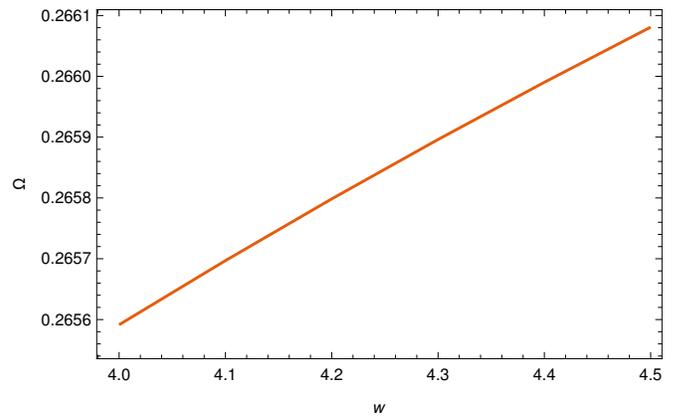}
\caption{$\Omega$ vs $\omega$ for lowest order ($n=0$) vortex solutions for first values of $\alpha$ that extremize $\Omega_{\alpha}$ in eigenvalue equation obtained by eq.(\ref{time dep. eom 1}). (see Appendix)}
\end{figure}

\section{Conclusion and Remarks}

\noindent In this work, we have holographically devised vortex solutions with different winding numbers in a rotating superfluid. These solutions may be interpreted as vortices placed at the centre of the disc at $r=0$. Our analysis shows that $p = \Omega R^2$ is an exact condition for $n=0$ case while it is true for $p >> 2$ for higher order solutions, that is, $n \neq 0$.
This linear relation between the winding number, $p$, and the angular velocity, $\Omega$, seems to be an universal feature of such vortices at least for large $p$. It is to be noted that due to the Neumann boundary condition at $r=0$, the vortex solution with winding number $p = 1$ is absent in this model. However, if one considers the Dirichlet boundary condition, at $r = 0$, instead of Neumann boundary condition, then even solutions with winding number $p = 1$ are allowed. In Fig.(1), we have shown some vortex solutions with different winding numbers. We have further solved the bulk equation using St{\"u}rm-Liouville eigenvalue approach and have observed that the chemical potential must be purely imaginary. A relation between winding numbers associated with the vortices and the imaginary chemical potential for the specific case of lowest order ($n=0$) vortices have been found. We have given a novel interpretation to this relation in terms of reduction in the number of vortices in rotating holographic superfluids, with increase in the imaginary chemical potential, which in turn implies reduction of the dissipation in the system. As a final remark, we would like to emphasize that the results in this work have been obtained analytically making use of the gauge/gravity duality and has similar features to those found numerically.

\noindent {\bf{Acknowledgements}}:  AS would like to thank Neeraj Kumar for some fruitful discussions. AS would also like to acknowledge anonymous referee for critical comments, which led us to some interesting findings.


\section{Appendix}

If we consider $\mu = 0$ in eq.(\ref{time dep. eom}), then we get
\begin{eqnarray}
 (1-u^3)\partial_u^{2}\Phi-\Big(u^2+\dfrac{2}{u}+2i\omega\Big)\partial_u \Phi  \nonumber \\-\Big(-\dfrac{2}{u^2}-\dfrac{2i\omega}{u}\Big) \Phi = \lambda \Phi ~.
\label{time dep. eom 1}
\end{eqnarray}
Now comparing eq.(\ref{time dep. eom 1}) with eq.(\ref{bulk equation 02}), we find that these two equations are similar to each other with the difference of sign in $\mu$ and $\omega$. Now using the St{\"u}rm-Liouville eigenvalue approach, we can solve eq.(\ref{time dep. eom 1}). The resulting behaviour between $\Omega$ and $\omega$ is given in Fig.(4). This figure shows that vortex number increases with the increase of quasi-normal frequency, $\omega$. This implies that dissipation of the system increases with increase in $\omega$.

\begin{thebibliography}{99}

\bibitem{prdR}Chuan-Yin Xia et.al., \href{https://link.aps.org/doi/10.1103/PhysRevD.100.061901}{Phys. Rev. D 100, 061901(R) (2019)}


\bibitem{cph}C. P. Herzog, \href{https://doi.org/10.1088/1751-8113/42/34/343001}{2009 J. Phys. A: Math. Theor. 42 343001}

\bibitem{rgc}R. Cai, L. Li  et al. \href{https://doi.org/10.1007/s11433-015-5676-5}{Sci. China Phys. Mech. Astron. 58, 1–46 (2015)}


\bibitem{hhh}S. A. Hartnoll, C. P. Herzog, G. T. Horowitz, \href{https://doi.org/10.1103/PhysRevLett.101.031601}{Phys. Rev. Lett. 101, 031601.}

\bibitem{hhh1} S. A. Hartnoll, C. P. Herzog, G. T. Horowitz, \href{https://doi.org/10.1088/1126-6708/2008/12/015}{JHEP12(2008)015}


\bibitem{sah}S. A. Hartnoll, \href{https://doi.org/10.1088/0264-9381/26/22/224002}{2009 Class. Quantum Grav. 26 224002}

\bibitem{gr1}S. Gangopadhyay, D. Roychowdhury, \href{https://doi.org/10.1007/JHEP05(2012)002}{J. High Energ. Phys. (2012) 2012: 2.}

\bibitem{gr2}S. Gangopadhyay, D. Roychowdhury, \href{https://doi.org/10.1007/JHEP05(2012)156}{J. High Energ. Phys. (2012) 2012: 156.}


\bibitem{sg1}S. Gangopadhyay, \href{https://doi.org/10.1016/j.physletb.2013.06.027}{Physics Letters B
Volume 724, Issues 1–3, 9 July 2013, Pages 176-181}



\bibitem{gg1}D. Ghorai, S. Gangopadhyay, \href{https://doi.org/10.1140/epjc/s10052-016-4005-0}{Eur. Phys. J. C (2016) 76: 146.}

\bibitem{js} J. Sonner,  \href{https://doi.org/10.1103/PhysRevD.80.084031}{Phys. Rev. D 80, 084031}

\bibitem{assg}A. Srivastav, S. Gangopadhyay, \href{https://doi.org/10.1140/epjc/s10052-019-6834-0}{Eur. Phys. J. C 79, 340 (2019).}

\bibitem{asdgsg}A. Srivastav, D. Ghorai, S. Gangopadhyay, \href{https://doi.org/10.1140/epjc/s10052-020-7769-1}{Eur. Phys. J. C 80, 219 (2020).}

\bibitem{rbsg}R. Banerjee, S. Gangopadhyay et.al. \href{https://doi.org/10.1103/PhysRevD.87.104001}{Phys. Rev. D 87, 104001}


\bibitem{gg2}D. Ghorai, S. Gangopadhyay \href{https://doi.org/10.1140/epjc/s10052-016-4005-0}{Eur. Phys. J. C 76, 146 (2016)}


\bibitem{gg3}D. Ghorai, S. Gangopadhyay \href{https://iopscience.iop.org/article/10.1209/0295-5075/118/31001/meta}{EPL 118, 31001 (2017)}

\bibitem{pmchl} P.M. Chesler, H. Liu, A. Adam \href{https://www.science.org/lookup/doi/10.1126/science.1233529}{Science 341 (6144) 368}

\bibitem{mno}K. Maeda, M. Natsuume, T. Okamura, \href{https://link.aps.org/doi/10.1103/PhysRevD.81.026002}{Phys. Rev. D 81, 026002 (2010)}

\bibitem{prl}M. Montull, A. Pomarol, P. J. Silva \href{https://link.aps.org/doi/10.1103/PhysRevLett.103.091601}{Phys. Rev. Lett. 103, 091601 (2009)}

\bibitem{cvj1}T. Albash and C. V. Johnson, \href{https://iopscience.iop.org/article/10.1088/1126-6708/2008/09/121}{JHEP09 121 (2008)}

\bibitem{cvj2}T. Albash and C. V. Johnson,, \href{https://journals.aps.org/prd/abstract/10.1103/PhysRevD.80.126009}{Phys. Rev. D 80, 126009 (2009)}

\bibitem{cvj3}T. Albash and C. V. Johnson,, \href{https://arxiv.org/abs/0906.0519}{arXiv:0906.0519v1}

\bibitem{ovlet}O. V. Lounasmaa, E. Thuneberg \href{https://www.pnas.org/content/96/14/7760.abstract}{Proc. Natl. Acad. Sci. USA 96, 7760-7767 (1999)}

\bibitem{gev}G. E. Volovik \href{https://iopscience.iop.org/article/10.3367/UFNe.0185.201509h.0970}{Physics-Uspekhi 58 (9) 897-905 (2015)}

\bibitem{xtzh} X. Li, Y. Tian, H. Zhang, \href{https://doi.org/10.1007/JHEP02(2020)104}{J. High Energ. Phys. 2020, 104 (2020).}

\bibitem{rw} A. Roberge and N. Weiss \href{https://doi.org/10.1016/0550-3213(86)90582-1}{Nucl. Phys.
B275, 734 (1986)}


\bibitem{gasp} G. Aarts, S. P. Kumar, and J. Rafferty  \href{https://doi.org/10.1007/JHEP07(2010)056}{J. High Energy Phys. 07 (2010) 056.}


\bibitem{jr}J. Rafferty  \href{https://doi.org/10.1007/JHEP09(2011)087}{J. High
Energy Phys. 09 (2011) 087.}

\bibitem{ck}G. E. Cragg and A. K. Kerman\href{https://link.aps.org/doi/10.1103/PhysRevLett.94.190402}{Phys. Rev. Lett. 94, 190402}

\bibitem{kg}K. Ghoroku et.al.\href{https://link.aps.org/doi/10.1103/PhysRevD.102.046003}{Phys. Rev. D 102, 046003}

\end{thebibliography}
\end{document}